\documentclass[a4paper]{jpconf}
\usepackage{graphicx}
\begin{document}
\title{Inclusive and Exclusive Scatterings from Tensor Polarized Deuteron}

\author{Misak M. Sargsian}

\address{Department of Physics, Florida International University, Miami, FL 33199}

\ead{sargsian@fiu.edu}

\author{Mark I. Strikman}

\address{Department of Physics, Pennsylvania State University, University Park, PA  16802}

\ead{strikman@phys.psu.edu}

\begin{abstract}
The possibility of using a  tensor polarized deuteron target in electroproduction reactions creates new opportunities 
for studying different phenomena related to the  short-range  hadronic and nuclear physics.  
The use of tensor polarized 
deuteron allows to isolate  smaller than average inter-nucleon distances for the bound two-nucleon system.
In this report we consider several  of  high $Q^2$ reactions   which are  particularly sensitive to 
the short-range  two-nucleon configurations in the deuteron.  
The one is the relativistic dynamics of  electron-bound-nucleon scattering which can be studied in both 
inclusive and exclusive reactions, other is the strong final state interaction  in close proximity  of two 
nucleons that can be used as a sensitive probe for color-transparency phenomena.
 \end{abstract}

\section{Introduction}
The deuteron is the simplest nuclear system with the wave function strongly dominated by  $pn$ component. 
Thus, it  can be used for studies of many  aspects of $pn$ strong interaction.   One of such aspects 
is the studies of the $pn$ system at short distances where one hopes to gain access to 
 many fundamental issues of nuclear dynamics, such as relativistic description of nuclear structure, 
 the dynamics of the $NN$ repulsive core, role of the non-nucleonic degrees of freedom, 
 and hadron-quark transition  at very short distances. 
 
However one problem in realizing such a program of studies is that the deuteron is barley bound 
with the charge-rms radius of about $2.1~fm$. This fact 
in the momentum space is reflected  in the very steep momentum 
distribution of the unpolarized deuteron wave function with 
the strength concentrated predominantly at the small relative 
momenta in the $pn$ system,  (see the curve labelled as $\rho^{unp}$ in  Fig.1). 

\begin{figure}[htb]
\centering\includegraphics[width=9cm,height=7cm]{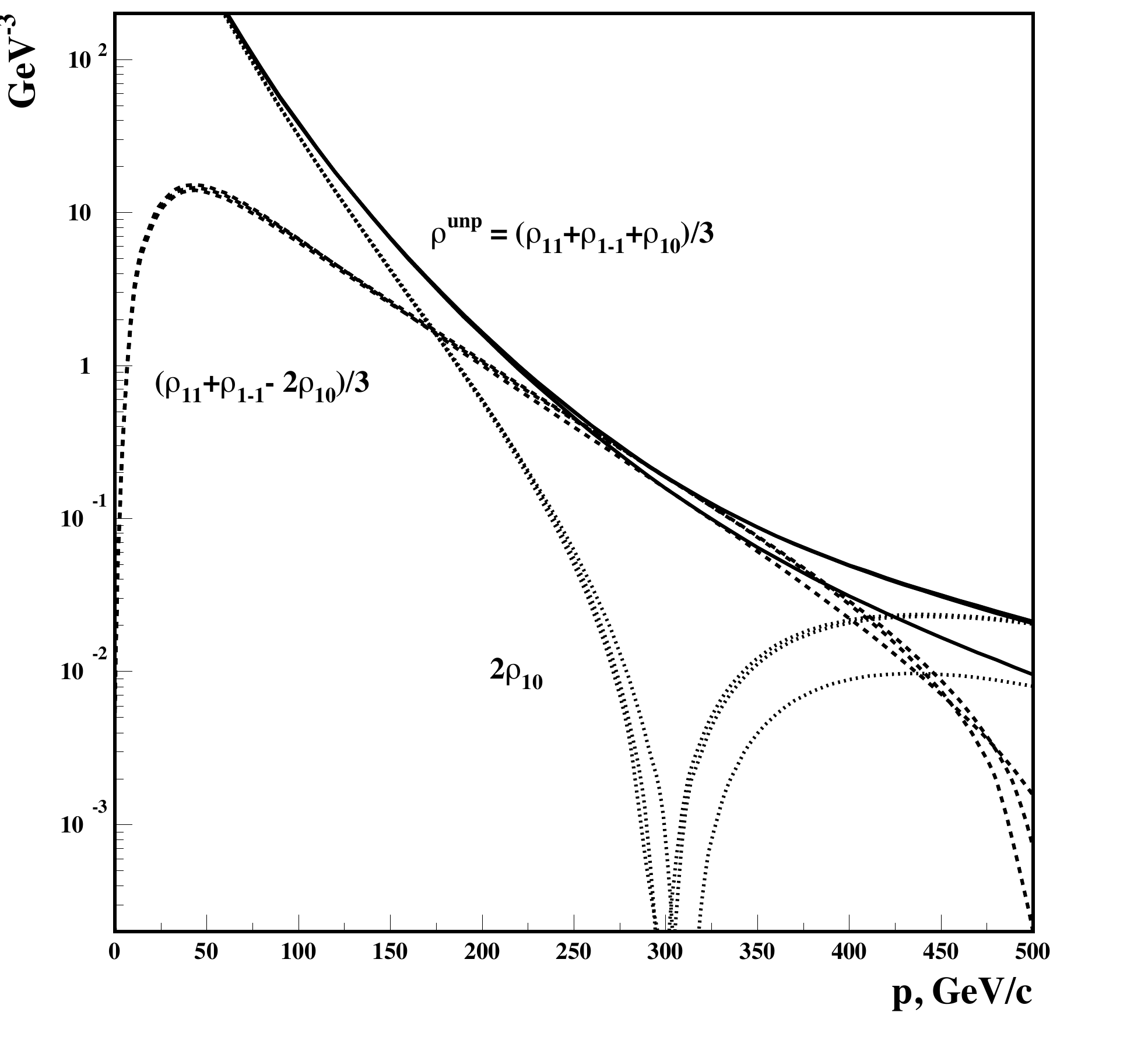}
\caption{Momentum dependences of different combinations 
of the polarized density matrices. Solid, dashed and doted curves correspond 
to  unpolarized, tensor polarized and transverse polarized  distributions 
respectively.}
\label{fig1}
\end{figure}

However the fact that the 
deuteron has a $D$-wave which vanishes at small momenta 
indicates that isolating it  in any 
given reaction with the deuteron  will allow effectively 
suppress the small-momentum/long-range contributions.

This can be seen from the polarized density matrix of the deuteron\cite{FGMSS95}:
\begin{eqnarray}
&&\rho_{d}^{\vec a}(k_1,k_2))    =   u(k_1)u(k_2) +
\left[1 - {3|k_2\cdot a|^2 \over k_2^2} \right]{u(k_1)w(k_2) \over \sqrt{2}} +
\left[1 - {3|k_1\cdot a|^2 \over k_1^2} \right]{u(k_2)w(k_1) \over \sqrt{2}}
\nonumber \\
\nonumber \\
&&+\left( {9\over 2} { (k_1 \cdot a)(k_2\cdot a)^*(k_1\cdot k_2)
 \over k_1^2k_2^2 }-
{3\over2}{|k_1\cdot a|^2\over k_1^2} - {3\over2}{|k_2\cdot a|^2\over k_2^2} +
{1\over 2}
\right) w(k_1)w(k_2),
\nonumber \\
\label{rho_a}
\end{eqnarray}
where $u(k)$ and $w(k)$ represent the $S$ and $D$ partial waves respectively.
The polarization vector  $\vec a$ is defined through the deuteron spin wave
functions:
\begin{equation}
\psi^{10} = i\cdot a_z , \  \psi^{11} = -{i\over \sqrt{2}}(a_x + ia_y) ,
                         \  \psi^{1-1} = {i\over \sqrt{2}}(a_x -ia_y),
\label{VK}
\end{equation}
where $\psi^{1\mu}$ is the projection of the deuteron's spin in the $\mu$
direction. The unpolarized density matrix of the deuteron is defined as:
$\rho^{unp}_d(k_1,k_2)  = {1\over 3} \sum_a  \rho_{d}^{a}(k_1,k_2)$.

Since $\lim_{k\rightarrow 0}w(k) =  0$, it follows from Eq.(\ref{rho_a}) that 
any polarization combination of $\rho_{d}^{\vec a}$,  in  which
$u^2$ term is canceled has an enhanced sensitivity to  the larger 
internal momenta (smaller distances)  of the  deuteron  as compared to the 
unpolarized case.  It follows from 
Eq.(\ref{rho_a}), that  the $u(k_1)u(k_2)$ term does not depend on the polarization 
vector $\vec a$, thus one can cancel this term summing any two polarization components 
of the density  matrix and subtracting the doubled value of the third  polarization
component. One example is:
\begin{eqnarray}
\rho^{20}(k_1,k_2) \equiv {1\over 3}(\rho^{11} + \rho^{1-1} - 2\rho^{10}) = 
\left({3k_{1z}^2\over k_1^2} - 1\right) {u(k_2)w(k_1)\over \sqrt{2}} + 
\left({3k_{2z}^2\over k_2^2} - 1\right) {u(k_1)w(k_2)\over \sqrt{2}}  \nonumber \\
+\left( {3\over 2}\left[{(k_1k_2)(k_1k_2 - 3k_{1z}k_{2z})\over k_1^2 k_2^2}
+{k_{1z}^2\over k_1^2} + {k_{2z}^2\over k_2^2}\right] - \right)w(k_1)w(k_2).
\label{rho_t20}
\end{eqnarray}

Fig.\ref{fig1} presents the  examples of the density matrices for 
unpolarized ($(\rho_{11} + \rho_{1-1} + \rho_{10})/3$,
transverse, $\rho_{10}$,  and tensor  polarized,  $\rho_{20}$, deuteron targets 
as they enter in the impulse approximation term of the  $\vec d(e,e'p)n$ 
cross section (in this case $k_1=k_2=p$). 
As it can be seen from Eq.(\ref{rho_t20}) the tensor polarized density matrix
depends only  on the terms proportional to $u(p)w(p)$ and $w(p)^2$.

This suggests the ways for studying  several issues of nuclear physics related to 
the short-range  interactions  using tensor polarized  deuteron targets.

\medskip

{\bf The $D$-wave Component of the Deuteron:} The presence of the $D$-wave component in the 
deuteron follows from the existence of the finite quadruple momentum of the deuteron. It is directly 
related to the tensor part of the $NN$ interaction.  The modern $NN$ interaction potentials which
fit the existing $NN$ phase-shifts  with $\xi^2\le 1$ predict  the overall probability of the $D$ wave 
in the deuteron to be in the range of  $4.87\%$ (cdBonn\cite{cdBonn}) to $5.76\%$ (AV18\cite{AV18}). 
Although the difference seems rather 
small, these parameterizations predict substantially different high momentum strengths above $400$~MeV/c,  
which is predominantly related to the $D$ component.  
The need to understand the D-wave momentum distribution became more pressing 
recently,  due to the  observation of the 
strong dominance of $pn$ short range correlations (SRCs) in nuclei as compared to the $pp$ and $nn$ 
SRCs\cite{isosrc,Subedi:2008zz}, which is related to the tensor component of NN interaction in 
the SRC\cite{eheppn2,Schiavilla:2006xx}. The dominance of the tensor interaction in the high momentum 
component of the nuclear wave function may play important role in the dynamics of the asymmetric 
nuclei\cite{newprops,proa2}  with nontrivial implications for superdense nuclear matter and neutron stars\cite{srcrev,progsrc}.

Very recently another interesting aspect of the role of the $D$-wave was revealed in the observation of $k^{-4}$ scaling 
of the momentum distributions in the deuteron and nuclei\cite{contact} in the  momentum range of 300-600~MeV/c,
which show  intriguing similarities with the contact
observed in the two-component symmetric  atomic systems of fermions.

As it follows from Eq.(\ref{rho_t20}) the processes involving tensor-polarized deuteron target are directly 
related to the structure of the $D$-partial wave component in the deuteron.

\medskip

{\bf Relativistic Dynamics of the  NN Bound System:}
One of the important issues in studying of nuclear structure  at short distances is the need for a 
relativistic description of the bound system.  This is an important issue also in 
understanding the QCD medium effects with recent studies indicating that  parton distribution 
modifications  in nuclei are proportional to the high momentum component of nuclear wave function\cite{emc_src}.

The deuteron is the simplest bound system and naturally any self-consistent attempt  to understand the 
relativistic effects in the bound nuclear systems  should start with the deuteron. 
The issue of the relativistic description of the deuteron has long history with extensive research 
started already in late 1970's (see e.g. \cite{Frankfurt:1977vc,Buck:1979ff,FSRep81,FSRep88}).

The experimental studies of the relativistic effects in the unpolarized  deuteron  up to now included the large $Q^2$ elastic 
$ed$ scatterings\cite{Alexa:1998fe}.  However   due to complexities  in the reaction mechanism the relativistic 
effects were  difficult to isolate.

The processes involving tensor polarized deuteron target are expected to exhibit  enhanced sensitivity to the 
relativistic effects due to higher average momentum of the bound nucleon entering in the polarized 
nuclear density matrix as compared to the unpolarized case. Thus one can discuss possible high momentum 
transfer reactions off tensor polarized deuteron and study their sensitivity to the relativistic effects of the 
bound nucleon motion.

{\bf Final Sate Hadronic Interactions:} The fact that tensor polarized deuteron is characterized by a  larger average 
internal momenta indicates that the two nucleons are in close proximity and scattering from such 
target can yield large final state hadronic interactions in exclusive reactions.  The latter can be used in studies of 
the hadronic properties of produced particles such as nucleons\cite{FGMSS95,FSS97,SM01} or 
vector mesons\cite{vmeson1,vmeson2,jpsi}.

\begin{figure}[ht]
\centering\includegraphics[width=10cm,height=7cm]{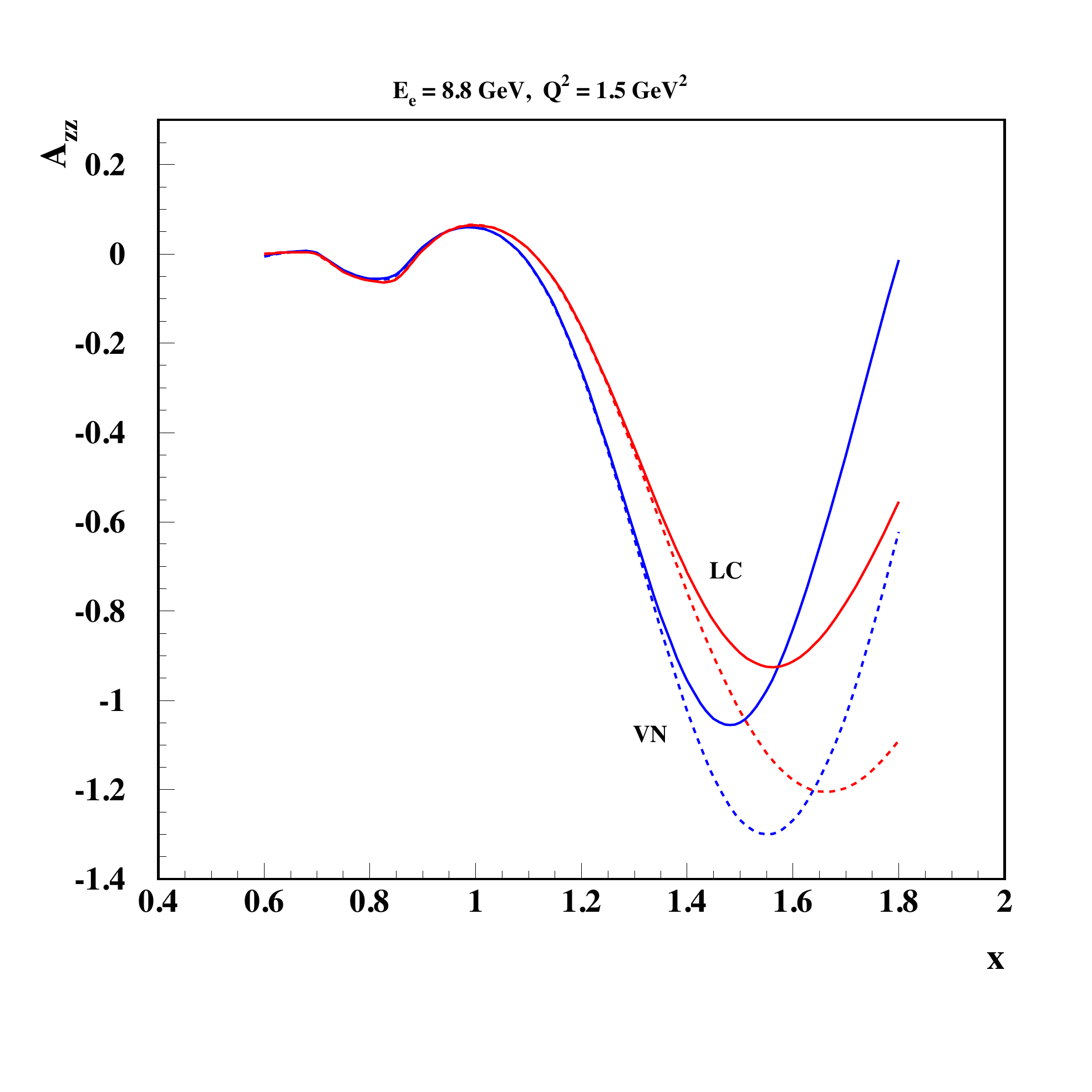}
\caption{The $x_{Bj}$ dependence of $A_{zz}$ asymmetry. The "VN"  and "LC" labels  correspond to 
the VNA and  and LC calculations described in the text.  Solid line  AV18(\cite{AV18})  and dashed line cdBonn (\cite{cdBonn}) 
NN potentials used in the calculations.}
\label{fig2}
\end{figure}

\section{Electroproduction Reactions Involving Tensor Polarized Deuteron}

We will discuss several electroproduction reactions in which utilizing the unique features of 
the tensor polarized deuteron would allow  to study the above discussed properties of high energy processes.

{\bf High $Q^2$ Near-Threshold Inclusive Scattering:}
High $Q^2$  ($>1$~GeV$^2$) inclusive scattering at $x_{Bj}>1$ is known to probe large longitudinal momenta of  bound nucleon 
interacting with the virtual photon.  One of the important indications that these reactions indeed  probe the high momentum 
component of  the nuclear wave function  is the observation of the scaling in the ratios of the $A(e,e^\prime)X$   cross 
sections to that of  deuteron\cite{FSDS,Fomin} or light nuclei\cite{Kim1,Kim2}. 

Using the tensor polarized
deuteron in such reactions allows to prepare the nucleus in the most compact state in which due to the absence of the 
pure (S-wave)$^2$  contribution (Fig.\ref{fig1})  the system in average is sensitive to a higher momentum of the nucleon 
in the deuteron for given $x$ and $Q^2$.
At large $Q^2>1$~GeV$^2$ kinematics the probed longitudinal momenta of the bound nucleon is $p_z \approx m_N(1-x)$, 
or the light cone momentum fraction $\alpha \ge x$. Because of these kinematic conditions and the absence of the 
 (S-wave)$^2$ contribution one expects a measurable relativistic effects already at $x\sim 1.2$.  Such an early onset 
of the relativistic effects indicates that one should be able to separate them from the uncertainty related  to the
choice of the  NN potential.

The sensitivity to the relativistic effects is estimated using the theoretical calculations based on two 
very different approaches.   The first approach describes the bound nucleon  in the 
deuteron rest frame  treating the interacting nucleon as being 
virtual (virtual nucleon approximation (VNA))  by taking the residue over the positive  energy pole of the spectator nucleon.
In this case the deuteron wave function satisfies the covariant equation of two-nucleon bound system 
with spectator being on mass-shell (see e.g. \cite{noredepn,Gross:2010qm}).

Another approach is based on the observation that high energy processes
evolve along the light-cone.  Therefore, it is natural to describe the 
reaction within the light-cone non-covariant framework \cite{FSRep81,FSRep88}. 
Negative energy states do not enter in this case, though one has to take into 
account so called instantaneous interactions.
In the approximation when non-nucleonic degrees of freedom in the
deuteron wave function can be neglected, one can relate
the light-cone wave functions to those calculated in the lab  frame
by introducing the LC $pn$ relative three momentum
$k=\sqrt{{m^2+p_t^2\over \alpha(2-\alpha)} - m^2}$.

In Fig.\ref{fig2}  the predictions for VNA\cite{noredepn} and LC\cite{FSDS} approximations are given 
for the tensor asymmetry,  $A^{zz}= {T^{20}\over \sigma^{unp}}$,
at  the highest $Q^2$ kinematics  for  the proposed recently experiment at Jefferson Lab\cite{loi}. As the figure shows 
for $1.2 \le x \le 1.4$ the uncertainty due to the  NN potentials used in the calculations is much smaller than the 
relativistic effects, while at high $x> 1.4$ we have substantial uncertainty due to the potentials.
The calculations predict large  $A_{zz}$ asymmetry in $x>1$ region with significant potential of 
discrimination  between both relativistic and NN potential effects.

{\bf High Momentum Transfer Exclusive Electrodisintegration of the Deuteron:}
The possibility of  the extension of inclusive measurements to the exclusive electrodisintegration reactions  in which 
struck nucleon carries almost all the momentum of the virtual nucleon gives direct access to the dynamics of the 
bound nucleon.  Recently the first high $Q^2$ experiments were completed for $d(e,e'p)n$ reactions\cite{Boeglin}, in which 
it was observed that due to the onset of the eikonal regime in the final state interaction (FSI)  of the stuck proton with 
the spectator neutron  it is possible to isolate kinematic regions with minimal and maximal FSI effects.  This observation 
is in agreement with the theoretical calculations which are based on high energy approximations in the description of  FSI 
effects (see e.g. \cite{FGMSS95,FSS97,SM01,Jesch,noredepn,Ford}).  

The pattern of FSI in $d(e,e^\prime p)n$ reaction is best seen by considering 
the  ratio of the cross section of the  full calculation to that of the plane wave impulse approximation:
$T = {\sigma^{Full}\over \sigma^{PWIA}}$. 
\begin{figure}[htb]  
\vspace{-0.4cm}  
\centering\includegraphics[width=10cm,height=7cm]{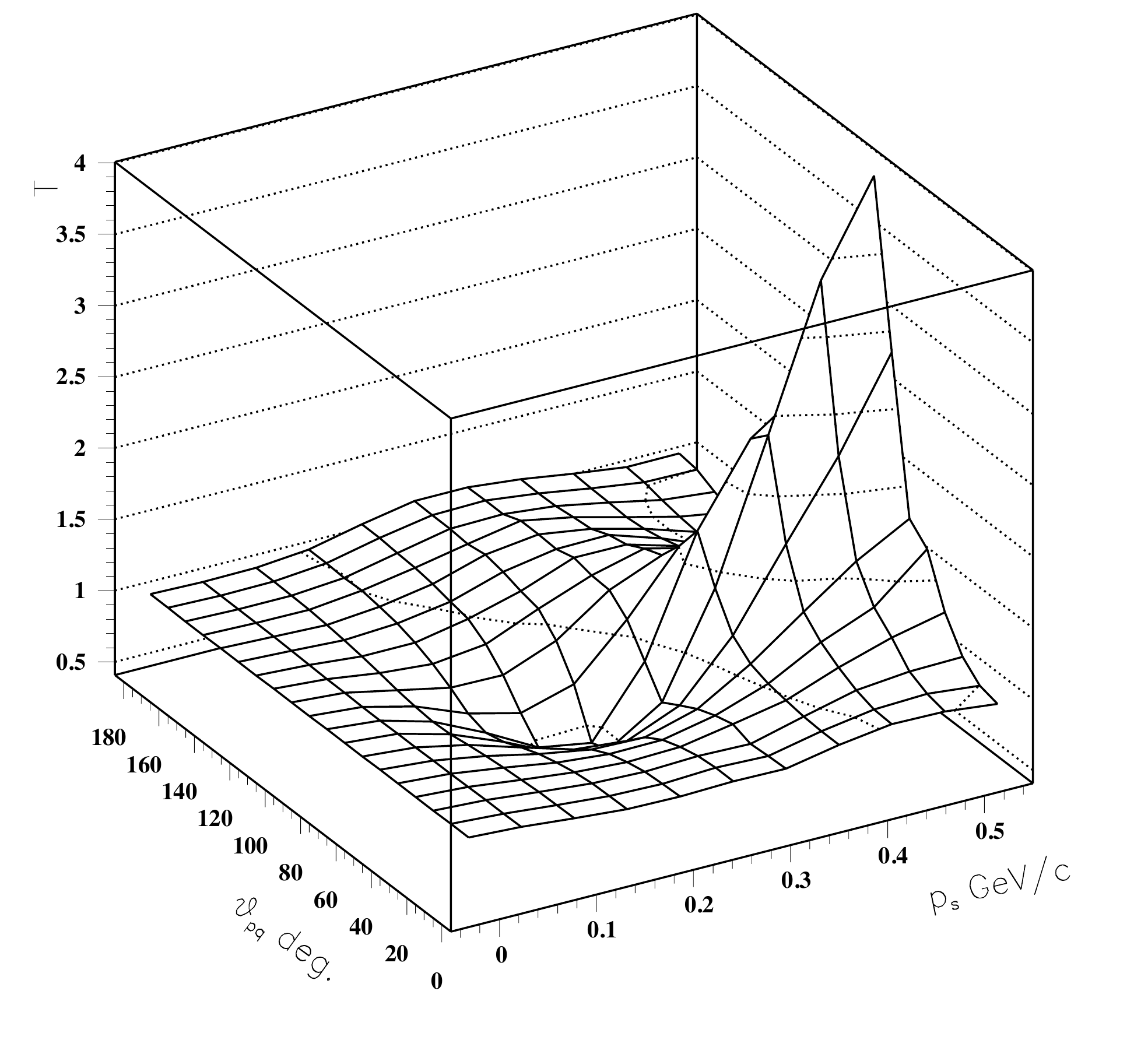}
\caption{The dependence of the transparency T on the angle - $\theta_{sq}$  
and the  momentum, $p_s$ of the recoil nucleon. The angle is defined relative 
to  ${\bf q}$.} 
\label{fig3}  
\end{figure}

Fig.\ref{fig3}   presents the  expectations for $T$ as a function of the  
recoil nucleon angle $\theta_{sq}$ relative  to  ${\vec q}$ for different 
values of recoil nucleon momentum. The figure demonstrates the distinctive angular 
dependence of the ratio $T$. At recoil nucleon momenta $p_s\le 300 MeV/c$,  
$T\le 1$  and has a minimum, while  at $p_s >300 MeV/c$, $T >1$ and  has a 
distinctive maximum. It can be seen from this  picture that the FSI 
is small at kinematics in which recoil momenta in the reaction is parallel or
anti-parallel to $\bf q$ (referred to as collinear kinematics).
The FSI dominates in the kinematics where  $\theta_{pq}\approx 90^0$, 
more precisely the maximal re-scattering corresponds to the kinematics in which 
$\alpha\equiv {E_s-p_{sz} \over m} = 1$ (referred to as transverse kinematics).
The analysis of Fig.\ref{fig3} shows that one indeed can isolate 
the kinematic domains where PWIA term is dominant from the domain in which 
FSI plays a major role. 
The ability to identify these two kinematics is an important advantage of 
high $Q^2$ $d(e,e^\prime p)n$ reactions. It allows to concentrate on the different aspects of 
the dynamics of $d(ee'p)n$ reaction with less background effects. Namely the collinear
kinematics are best suited for studies of bound nucleon dynamics while in transverse 
kinematics one can concentrate on the physics of hadronic re-interaction.

\begin{figure}[htb]
\vspace{-0.4cm}  
\centering\includegraphics[width=10cm,height=7cm]{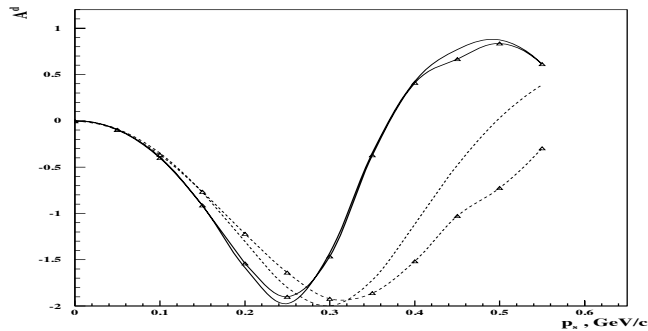}
\vspace{-0.4cm}  
\caption{The $p_s$ dependence of the $d(e,e'p)n$
tensor polarization asymmetry at $\theta_{sq}=180^0$. Solid
and dashed lines are PWIA predictions
of the LC and VNA approximations. The marked curves  present the  same (LC, VNA)  calculations including  
the FSI effects.}
\label{fig4}
\end{figure}

To quantify these statements in Fig.\ref{fig4} we present the calculation of $A_{d}$ (which is similar to $A_{zz}$ but for 
exclusive processes) in collinear  kinematics ($\theta_{sq}=180^o$)  within virtual nucleon and 
light-cone approximations described above.  These estimates, similar to the inclusive scattering case, show significant 
relativistic effects already at moderate momenta of $250-300$MeV/c for spectator nucleon.
It is interesting that the  FSI effects  further increase  the difference between VNA and LC predictions,
making it even easier to discriminate between VNA and LC approximations.

{\bf Dynamics of the Final State Interactions:}
As it follows from Fig.3, studying the $d(e,e^\prime p)n$ reactions at transverse kinematics  allows us to enhance the effects 
due final state interaction of struck nucleon off the spectator nucleon.  One expects that FSI effects will be further 
enhanced for tensor-polarized deuteron since it corresponds, in average,  to a smaller configuration than 
in the case of the unpolarized target.

\begin{figure}[htb]
\centering\includegraphics[width=10cm,height=6cm]{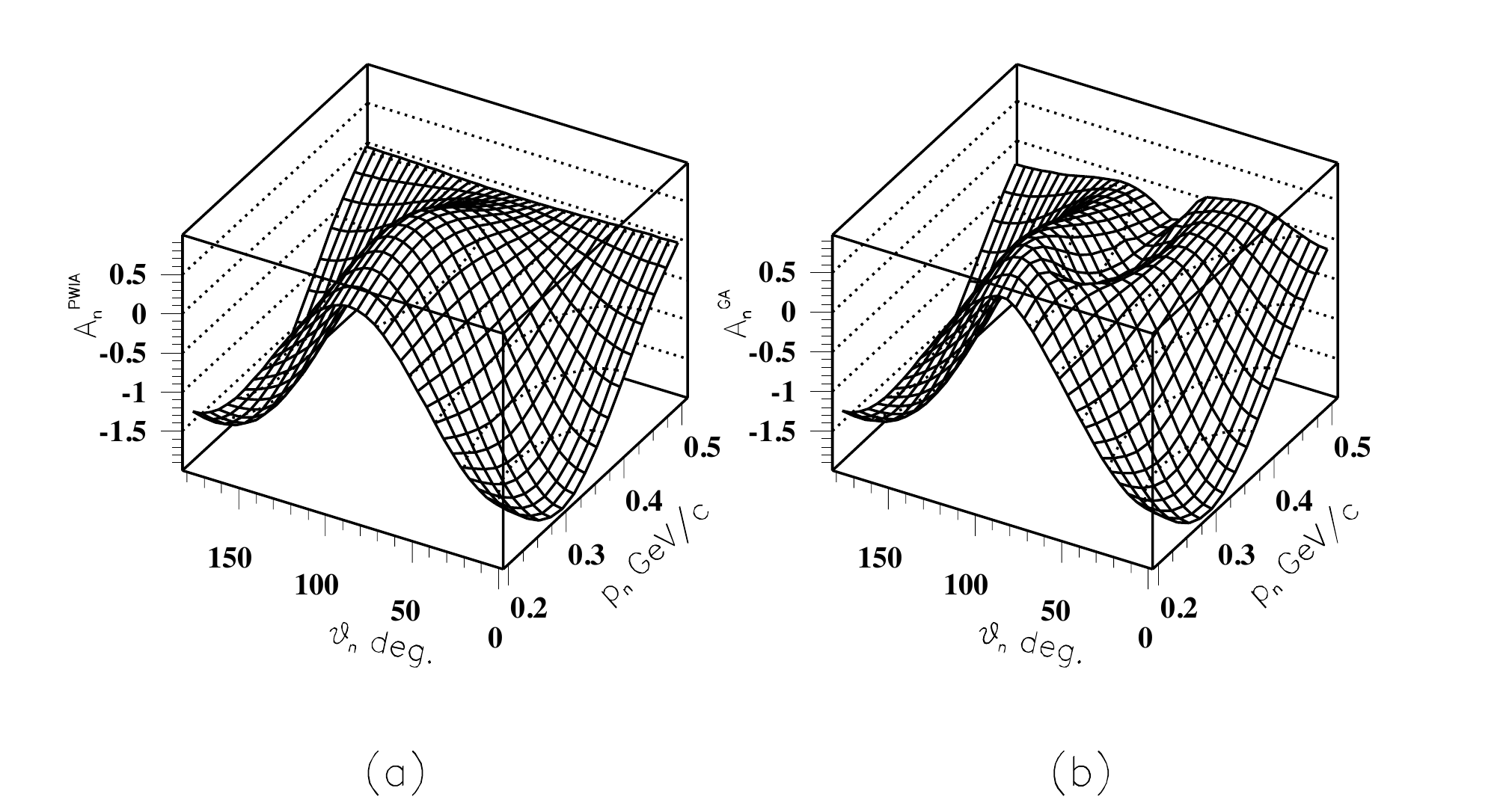}
\caption{The dependence of tensor asymmetry $A_d$ on the angle and momentum of the 
spectator neutron  in $d(e,e^\prime p)n$ reaction.  (a) PWIA  approximation (b) Full calculation 
that includes FSI.}
\label{fig5}
\end{figure}

If one assumes that FSI is determined by diffractive small angle $pn$  rescatterings one can estimate  the effect of the 
FSI on $A_d$, which is presented in Fig.\ref{fig5}.

The comparison of  Fig.\ref{fig5}(a) and (b) shows the very strong sensitivity of the asymmetry $A_d$ to the final state interaction at 
the transverse kinematics: the FSI significantly diminishes the asymmetry starting at $p_n = 300$~MeV/c and 
$\theta_{nq}\sim 90^0$ (or $\alpha_n \approx 1$).   This 
indicates that the quantity $A_d$ can be used as a very sensitive observable for studying the FSI dynamics: for example 
the onset of the Color Transparency~(CT)  phenomena (for details of CT phenomena see Ref.\cite{ANN}) 
 at large $Q^2$ ($\ge 4$~GeV$^2$).

\begin{figure}[htb]
\centering\includegraphics[width=10cm,height=7cm]{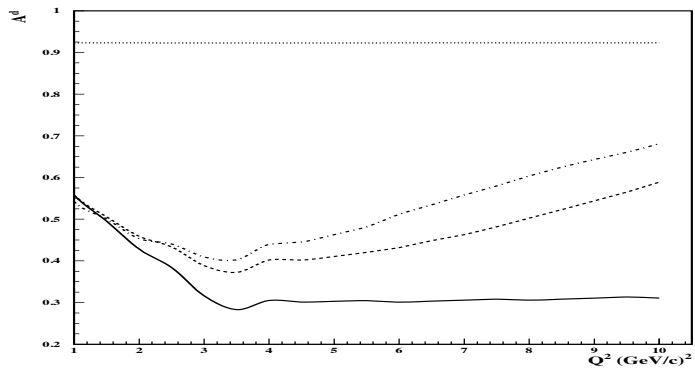}
\vspace{-0.5cm}
\caption{The $Q^2$ dependence of $A_d$ for $\alpha_n =1$. 
Solid line - Full calculation without CT effects, dashed  prediction of  Quantum Diffusion Model\cite{FLFS},
dashed-dotted - prediction of three state  resonance model model\cite{FGMS93}, dotted -PWIA.}
\label{fig6}
\end{figure}

For numerical estimates, we consider the $Q^2$ dependence of the asymmetry $A_d$ 
for fixed and transverse momenta of the spectator neutron.
This dependence for $p_t=300$ MeV/c, is presented in Fig.\ref{fig6}. 
One can see from this figure that CT effects  
can change $A_d$ by as much as factor of two for $Q^2\sim 10$ GeV$^2$.
It is worth noting that the same models predict  only 10-15\% effect for 
$(e,e'p)$ reactions on unpolarized nuclear targets.

\section{Conclusion and Outlook}
The use of the tensor polarized deuteron targets in high energy electro-production reactions
would provide a unique possibility for  studying the  NN strong interaction at short space-time separations.
We demonstrate that such reactions allow a direct test of the relativistic 
effects in the short range NN interaction. They also allow to discriminate  between different
NN potentials. 

Another advantage of the  tensor polarized deuteron as  a compact two-nucleon system,  is the possibility 
to enhance significantly the final state hadronic interaction in the exclusive $d(e,e^\prime,p)n$ reactions.
This provides 
a very sensitive tool for studying color transparency phenomena which will 
be manifested in the suppression of the FSI effects at large $Q^2$ kinematics.

All these studies are a part of the broad program dedicated to the investigation 
of the structure of deeply bound nucleon as well as the physics of color transparency\cite{hnm}.
This program could benefit tremendously from the advances 
of building polarized deuteron targets that can be operated with high 
current electron beams.

\medskip
 
\noindent {\bf Acknowledgements}\\
This work is supported  by  United States DOE grants under contract DE-FG02-01ER-41172 and DE-FG02-93ER40771.
\\

 \end{document}